# Plug&play fibre-coupled 73 kHz single-photon source operating in the telecom O-band

Anna Musiał[1,*], Kinga Żołnacz[2], Nicole Srocka[3], Oleh Kravets[1], Jan Große[3], Jacek Olszewski[2], Krzysztof Poturaj[4], Grzegorz Wójcik[4], Paweł Mergo[4], Kamil Dybka[5], Mariusz Dyrkacz[5], Michał Dłubek[5], Kristian Lauritsen[6], Andreas Bülter[6], Philipp-Immanuel Schneider[7], Lin Zschiedrich[7], Sven Burger[7,8], Sven Rodt[3], Wacław Urbańczyk[2], Grzegorz Sęk[1], and Stephan Reitzenstein[3,*]

[1]*Laboratory for Optical Spectroscopy of Nanostructures, Department of Experimental Physics, Faculty of Fundamental Problems of Technology, Wrocław University of Science and Technology, Wybrzeże Wyspiańskiego 27, 50-370 Wrocław, Poland*

[2]*Department of Optics and Photonics, Faculty of Fundamental Problems of Technology, Wroclaw University of Science and Technology, Wybrzeże Wyspiańskiego 27, 50-370 Wroclaw, Poland*

[3]*Institute of Solid State Physics, Technische Universität Berlin, Hardenbergstraße 36, 10623 Berlin, Germany*

[4]*Laboratory of Optical Fibers Technology, Institute of Chemical Sciences, Faculty of Chemistry, Maria Curie Sklodowska University, Maria Curie Sklodowska Sq 3, 20-031 Lublin, Poland*

[5]*Fibrain Sp. z o.o., Zaczernie 190F 36-062 Zaczernie, Poland*

[6]*PicoQuant GmbH, Rudower Chaussee 29, 12489 Berlin, Germany*

[7]*JCMwave GmbH, Bolivarallee 22, 14050 Berlin, Germany*

[8]*Zuse-Institute Berlin, Takustraße 7, 14195 Berlin, Germany*

[*] Corresponding authors: anna.musial@pwr.edu.pl; stephan.reitzenstein@physik.tu-berlin.de

**ABSTRACT**

A user-friendly fibre-coupled single-photon source operating at telecom wavelengths is a key component of photonic quantum networks providing long-haul ultra-secure data exchange. To take full advantage of quantum-mechanical data protection and to maximize the transmission rate and distance, a true quantum source providing single-photons on demand is highly desirable. We tackle this great challenge by developing a ready to use semiconductor quantum dot (QD)-based device that launches single photons at a wavelength of 1.3 µm directly into a single-mode optical fibre. In our approach the QD is deterministically integrated into a nanophotonic structure to ensure efficient on-chip coupling into a fibre. The whole arrangement is integrated into a 19" compatible housing to enable stand-alone operation by cooling via a compact Stirling cryocooler. The realized source delivers single photons with multiphoton events probability as low as 0.15 and single-photon emission rate up to 73 kHz into a standard telecom single-mode fibre.

**INTRODUCTION**

Sources of single photons (SPSs) are fundamental building blocks for photonic quantum technology, e.g., secure quantum communication[1,2], quantum internet[3], and linear quantum computation[4,5]. Recent world-wide activities on the implementation of quantum networks[6–11], including a satellite node[12], reflect the importance of the field. Among different SPS concepts, to date the purest single-photon emission is provided by semiconductor quantum dots (QDs) featuring probabilities of multiphoton events as low as $10^{-5}$ for emission wavelengths below 1 μm[13] (excluding their use for long-haul transmission[14–16]), and $10^{-4}$ at telecom wavelengths under non-resonant excitation[17]. Thus QDs constitute superb quantum emitters in terms of scalability, integration and compatibility with advanced semiconductor technology[18–25]. A general drawback hindering real-world applications of In(Ga)As QDs is the cryogenic operation temperature. In fact, this is the main reason why commercially available QKD systems and experimental quantum networks are almost exclusively based on sources utilizing spontaneous parametric down conversion or attenuated lasers[7–12,26–28]. This is despite the drawbacks that the former is probabilistic with low efficiency, whereas the latter does not inherently provide single photons making transmission susceptible to the photon number splitting attack[29].

In this work, we focus on developing a user-friendly SPS for quantum communication in the telecom O-band. This spectral window features a local minimum of loss in silica fibres, zero dispersion and is suitable for multiplexing with C-band classical signals, without the need for expensive dark fibres for the quantum channel, due to good spectral isolation reducing Raman scattering. Interestingly, fibre-based quantum links have already been implemented using QDs at the telecom O-band by KTH Stockholm and by Toshiba-Cambridge[30]. However, in all these reports the source was operated in complex and bulky experimental setups with QD emission coupled externally to a fibre. In order to take the application of QDs in quantum technologies to the next level we developed a user-friendly "plug&play" QD SPS which is fibre-coupled, compact and portable, includes a cooling system and provides a stable train of spectrally filtered single photons in the O-band via a standard telecommunication single-mode fibre. Importantly, this source is very convenient for the end user as it does not require any adjustment and is fully operational after a 15-minute cooldown cycle. In contrast, commercially available QD-based sources[31] utilize a standard bulky and expensive experimental cryostat and proof-of-principle realizations of compact designs[32] are multimode fibre-coupled. Moreover, both approaches operate at shorter wavelengths below 1 µm and require external spectral filtering of the single quantum transition. A previous work reporting on a plug&play-like SPS operating in the telecom O-band[14] is based on the random positioning of a single-mode fibre bundle with respect to the regular non-deterministic micropillar array featuring 30% multiphoton events. Whereas in our case, the QD-mesa fabrication technology and the positioning of the fibre with respect to the mesa are fully deterministic, which allows for the utilization of a single-mode fibre and the integration with a spectral filtering system. Overall, this is far beyond what has been reported so far in the field.

**RESULTS**

**Approach and design**

Our concept for realization of a compact fibre-coupled single-photon source is presented in Fig. 1a). The overall scheme aims at providing the end user with a stable source of internally or externally triggered 1.3 µm single photons directly in a standard single-mode fibre. The source is easy to handle and operates at stable photon flux without the need of any alignment or additional spectral filtering, and can be further used, e.g., for implementing quantum communication or computation schemes.

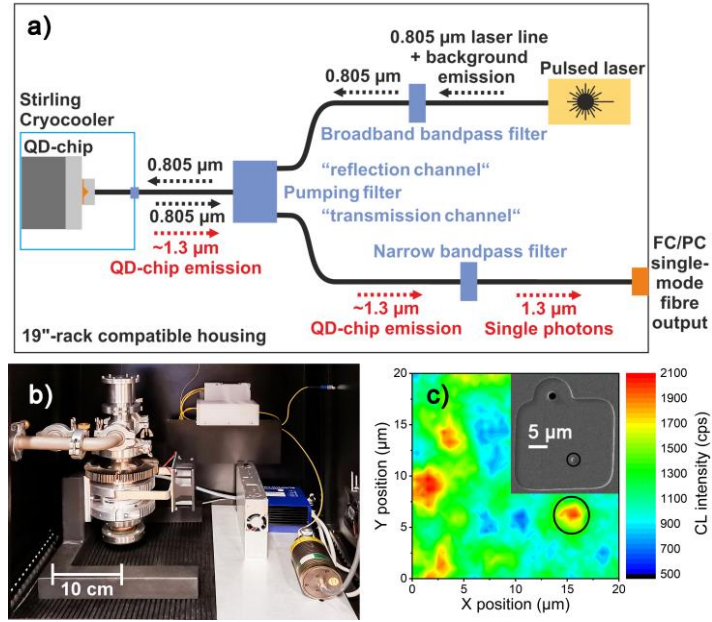

Fig. 1. a) Scheme of the fully fibre-coupled single-photon source - the frame marks the device with a standard telecom single-mode FC/PC fibre connector as output. b) Image of the actual device with Stirling cryocooler on the left, pulsed excitation laser on the bottom-right and the fibre components secured on the back wall. The ventilation openings in the housing and extra fans provide air flow for cooling of the Stirling cryocooler, which is operated in vertical position and mounted to the metal frame. c) Active region and sample patterning: spatial low-temperature (15 K) cathodoluminescence map in the spectral range corresponding to the target wavelength (1293-1295) nm with the emission intensity color-coded and the target QD marked by a black circle. Inset: scanning electron microscopy image of the patterned sample with mesa etched at the position of the target QD marked by a black circle.

The frame in the scheme marks elements which are placed in a 19" compatible housing shown in the image in Fig. 1b). The desired functionality is realized in the following way: Emission of the pulsed (80 MHz, pulse length <50 ps) non-resonant (805 nm) fibre-coupled electrically-triggered semiconductor diode laser is first spectrally filtered to transmit only the laser line itself into the fibre arrangement. The laser filter avoids unwanted broadband emission background, e.g., spontaneous emission from the laser cavity at the fibre output of our quantum device. In fact, while broad background emission (typically in the range of 600 – 1600 nm) from the laser is too low to efficiently excite QDs, it is comparable to the single QD signal. Therefore, it is crucial to filter this background out directly after the excitation laser. Once background emission passes the pumping filter in the all fibre-configuration it would not be filtered out by any other component and it would be spectrally integrated by the single-photon detectors into the photon flux. In that case, even the low spectrally-broad laser background could result in a total number of photons exceeding the number of photons from a single optical QD transition and therefore would make the whole system completely unusable. The spectrally filtered single-mode laser signal is delivered to the sample via a reflection channel (see Fig. 1a)) of a specially designed three-port filter for pumping. The aforementioned fibre components are based on standard telecom fibres and they are spliced to a high numerical aperture fibre (NA = 0.42), which is precisely positioned with respect to the single-photon emitter via a recently developed interferometric method[33] with an alignment accuracy of 50 nm and fixed to the sample surface by low temperature compatible epoxy glue (see Methods for details). It is important for a potential commercialisation of our concept to perform not only the fibre alignment in a deterministic way, but also the device fabrication itself. For that we apply in-situ electron beam lithography to pre-select suitable QDs via cathodoluminescence mapping before integrating them into microstructures with a process yield >90% and a positioning accuracy of about 40 nm[34]. This way, cylindrical mesas were deterministically fabricated around pre-selected QDs in approx. 20 µm x 20 µm writing fields (see inset to Fig. 1c)). In addition, the writing fields act as apertures and facilitate the orientation on the sample surface by mapping of its topography via a fibre in the alignment step. An exemplary CL map is shown in Fig. 1c) with the CL emission intensity color-coded for a narrow spectral range corresponding to the target wavelength of 1293-1295 nm at 40 K[34]. The position of the selected QD for device processing is marked by a black circle in Fig. 1c). It is characterized by emission in the spectral range of interest, high intensity and good spatial separation, which indicates that it originates from a single QD. This pre-selected QD is deterministically integrated into a cylindrical mesa with a diameter of (1090 ± 50) nm. Emission from this QD-micromesa is collected via the high-NA fibre and enters a transmission channel of the pumping filter which at the same time filters out the excitation signal. Finally, the target excitonic line of the QD is spectrally selected from emission of other QD transitions such as the biexciton via a narrow (0.6 nm) fibre-integrated bandpass filter. The exit port of this filter is connected to the FC/PC fibre connector at the optical output of our stand-alone single photon source. The sample design, the microstructure and fibre geometry, as well as the placement of the fibre follow the results of numerical optimization of the design parameters. To this end, the propagation of the light emitted by the QD was simulated using a finite-element method[35]. The system parameters with optimal fibre-coupling efficiency were determined using Bayesian optimization as global optimization method[36].

One of many technological challenges in implementing our device concept is the filtering of all spurious spectral contributions besides the single photons originating from the target QD transition from the all-fibre coupled system. This is realized by customized fibre components described in detail below. These optical elements have to provide high spectral isolation of the target QD transition with minimum insertion loss.

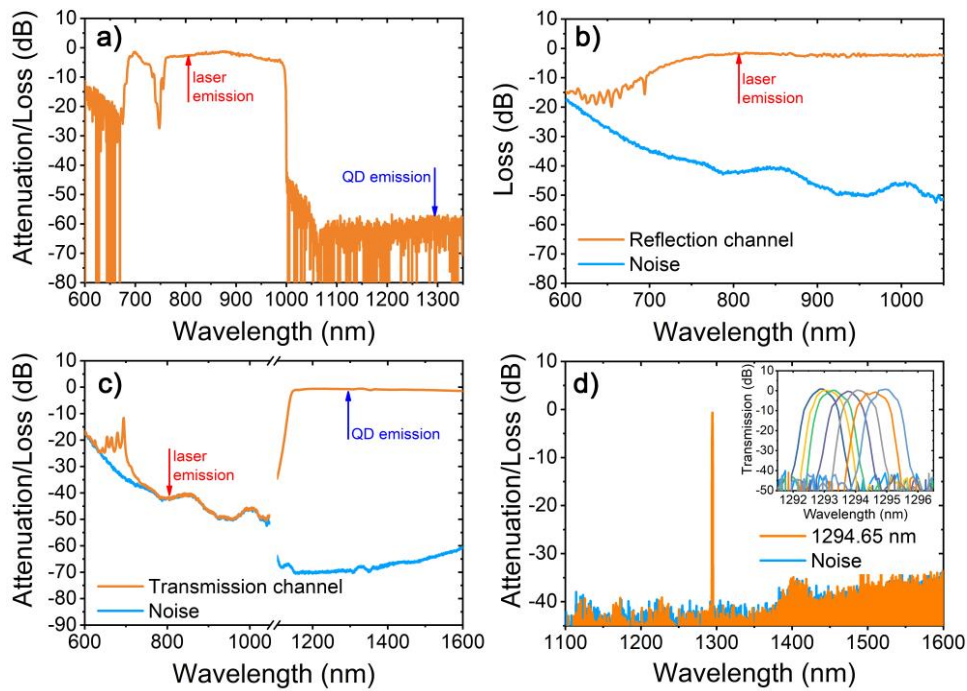

Fig. 2. Spectral characteristics of the fibre components measured with the supercontinuum and optical spectrum analyser (OSA): a) Broadband bandpass filter for the excitation laser. b) Single-mode (SM) pumping filter – reflection channel through which the optical excitation is delivered to the sample after cleaning up its spectrum with the filter presented in a). c) SM pumping filter – transmission channel through which the QD signal is delivered to the detection system. The laser radiation is blocked by -40 dB; d) SM narrow bandpass flat-top filter (0.6 nm full width at half maximum) with a central wavelength of 1294.65 nm; inset: set of fiber-based narrow bandpass differing in the adjustable central wavelength covering 2 nm spectral range of 1293 to 1295 nm. The orange curves show the transmission of the investigated component with respect to the reference signal level and the blue one depicts the OSA noise level.

**Specialty fibre components**

Customized fibre components were designed and fabricated to fulfil the requirements of the stand-alone SPS device concept. Their optical characteristics (see Fig. 2) were evaluated using a supercontinuum light source and an optical spectrum analyser (OSA). The laser filter responsible for cleaning preliminarily the laser spectrum (see Fig. 2a) has high transmission at the laser line wavelength (805 nm) of -2.5 dB and high attenuation in the spectral range of the QD emission (-62 dB). For the flexibility of the optical excitation, this filter is a broad bandpass to allow for application of different laser wavelengths in the range of (700 – 1000) nm. Laser light in this range is spectrally removed in the transmission channel (Fig. 2c) of the fibre pumping filter with an optical isolation better than -35 dB in the broad range and better than -40 dB at the actual laser line position. In the final device, the laser attenuation was increased to >80 dB by combining two filters of this type. The results of the measurements are shown for a single filter due to limited dynamic range of the OSA. This channel exhibits high transmission for the optical signal from the QD with -0.7 dB loss above 1150 nm up to at least 1700 nm (OSA detection limit) covering both the telecom O-band and C-band. In the reflection channel, the transmission is high in the range above 750 nm (loss lower than -4.2 dB) so that the optical excitation can be delivered efficiently to the sample. The relatively high loss at the wavelength of the laser line is not an issue in the case of single QDs as for the investigated structures the emission is typically saturated at average excitation powers in the single μW range (cf. Fig. 3b). The next step is to isolate a single optical transition from the other emission signals related to the wetting layer, strain reducing layer, possible defects or other excitonic complexes confined in the same QD. This functionality is realized by an ultra-narrow top-flat fibre-integrated bandpass filter (0.6 nm) (see Fig. 2d)). To provide some spectral, a set of 7 exchangeable filters covering a wavelength of 1293 to1295 nm in the O-band with isolation better than -45 dB was fabricated.

**Device performance**

The optical properties of the stand-alone telecom single-photon source are evaluated using a fibre based Hanbury Brown and Twiss configuration equipped with superconducting nanowire single-photon counting modules (SNSPDs). Figure 3a) depicts the corresponding coincidences' histogram (black curve) obtained under pulsed excitation at 80 MHz with an average excitation power of 0.75 μW recorded at the input of the customized fibre arrangement (T = 40 K), which corresponds to 0.65 μW incident on the sample. The measured histogram was fitted by the sum of double-sided exponential decays for each maximum[17] including the background level in between the emission pulses. The peak height for the non-zero delay peaks determined from the fitting procedure was further used to normalize the coincidence events histogram to obtain the time delay-dependent second order correlation function $g^{(2)}(\tau)$. The probability of multiphoton emission events $g^{(2)}(0)$ was determined from the fitting procedure as the ratio of the height of the central (zero delay) peak and the peaks at the long time delays and yields background-corrected $g^{(2)}(0) = 0.15\pm0.05$ proving single-photon emission from the target optical transition. Here, the level of uncorrelated background determined in between the emission pulses is subtracted from the as measured $g^{(2)}(0)$ (see Methods for details). Here, the uncorrelated background signal is mainly attributed to non-ideal laser suppression in the full-fiber configuration. This issue can most probably be resolved by further increasing the attenuation of the laser blocking filter in the future. The associated PL spectrum of the QD at the output of the SPS is shown in the inset of Fig. 3a). The emission is centred at 1294.7 nm, and the linewidth equals 0.43 nm which is a typical value for 1.3 µm QDs[18,37], where the quite significant inhomogeneous broadening is related mainly to spectral diffusion effects in the case of non-resonant excitation. At this excitation strength, the total photon flux yields 31 kHz at the device output which, taking into account the 15% probability of multiphoton events, corresponds to true single photon rate of 27 kHz – Fig. 3b), where the latter rate considers the number of multiphoton events according to Ref.[38]. These emission rates are calculated at the output of the demonstrator (after the narrow bandpass filter), so the actual count rates on the SNSPDs are divided by the measured setup efficiency. Therefore, this is the actual single-photon rate that the end user will be provided with.

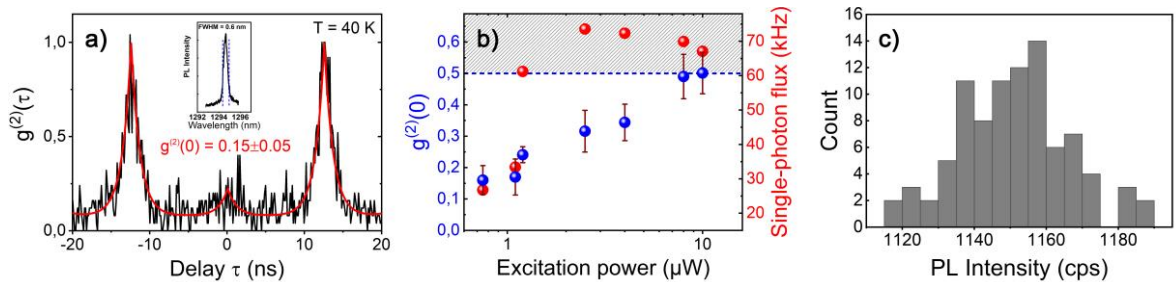

Fig 3. Optical properties of the stand-alone SPS: a) Normalized coincidences histogram measured under pulsed non-resonant excitation (0.75 µW average power at 80 MHz) in all-fibre configuration at T = 40 K (black curve) together with a fitting curve (red); inset: corresponding spectrum measured in all-fibre configuration including the fibre-based narrow bandpass filter (its bandwidth is marked with the dashed vertical lines); b) $g^{(2)}(0)$ (left scale, blue symbols) and single photon flux rate at the output of the demonstrator (right scale, red symbols), measured under pulsed non-resonant excitation at T = 40 K as a function of average excitation power; the red dashed line marks the $g^{(2)}(0) = 0.5$ defining the limiting value for single-photon operation; the error bars were obtained as the sum of the errors of fitting parameters; c) Stability test of the SPS: histogram of the count rates (averaged over 10 min) on one of the SNSPDs measured for 18 hours. A statistical analysis yields a mean value of 1151 counts per second (cps) and a standard deviation of 16 cps which corresponds to a 0.014 relative standard deviation.

To investigate the upper limit of the achievable single photon flux in the present device, coincidences histograms were measured as a function of the average excitation power in the range of 0.75 μW up to 10 μW. The corresponding power dependences of $g^{(2)}(0)$ – blue symbols and single-photon flux – red symbols, are presented in Fig. 3b). The limit of single-photon emission ($g^{(2)}(0) = 0.5$) is observed at 10 μW excitation power suggesting that at this excitation strength emission from the QD is already saturated, and that a further increase of the excitation power results in increased uncorrelated emission background overlapping spectrally with the QD line. The associated maximal true single photon flux corresponding to the saturation of QD emission equals 73 kHz. Its excitation power-dependence follows the emission intensity dependence of the single QD transition.

During collection of the histogram at a given excitation power the photon count rate at the SNSPDs was monitored over 18 h to get an insight into the long-term stability of the output of the source. The SNSPD detector count rates were averaged over 10 min and combined to generate the histogram presented in Fig. 3c). In comparison to emission rates in Fig. 3b this rate is decreased by the transmission of the experimental setup, including fibre beam splitter, fibre connectors and quantum efficiency of the detectors themselves. The statistical evaluation of these data yields a mean value of 1151 cps with a standard deviation of 16 cps, which corresponds to the relative standard deviation of 0.014. This shows that the long-term stability of the demonstrator output is better than 1.5%.

**DISCUSSION**

The realized plug&play SPS operates in the telecom O-band at a temperature of 40 K and provides a train of triggered single photons at a rate as high as 73 kHz into a single-mode fibre. Noteworthy, the rate of generation of single photons is by two orders of magnitude larger than the ~ 0.7 kHz reported in Ref.[32] for a multi-mode fibre-coupled stand-alone SPS based on a standard InGaAs QD emitting in 900-950 nm wavelength range. The lowest achieved probability of multiphoton events yields 15%. This highlights the significant advances achieved in the present work, which not only provides a fully fibre-based solution, but also demonstrates single-mode operation in the O-band – all of which are crucial prerequisites for real-world device applications, e.g., in the field of quantum communication. For optimized performance InGaAs QDs need to be cooled down. To take advantage of the compact and cheap cooling method provided by Stirling cryocoolers a direct, rigid and thermally as well as mechanically stable fibre coupling of the QD emission to a single-mode fibre was developed[33]. This overcomes the drawback of low frequency vibrations with the amplitude in the range of μm exceeding the size of the quantum emitter (at most tens of nanometers) inherent to the operation of the Stirling cryocooler. The proposed solution with superior performance is obtained by the interplay of various developed components, pump laser, growth of high-quality self-assembled QDs and design of mesa-DBR-fibre-setup based on numerical modelling, deterministic fabrication of QD-mesas, cooling method, specialty high-NA fibre, high-precision fibre positioning and customized fibre components for spectral filtering. An important challenge of our device concept is to deliver the excitation efficiently and filter out the unwanted photons both from the pump laser and emitted from other parts of the structure. High isolation of the single transition is needed both in broad range as well as in a narrow range. Due to relatively small binding energies (in the range of 1 meV) of various excitonic complexes confined in the same QD[37], a very sharp edge bandpass filter is required. Even more important is the loss for the actual single-photon signal. Commercially available elements can offer arbitrary good isolation, but the insertion loss, especially if one has to stack several of such elements with different functionalities, is typically unacceptable with values in the range of 1.5 dB per element for a signal from a single QD. Therefore, minimizing the insertion loss of the fibre components was crucial for the realization of the reported source. This can be further optimized by using low loss splices instead of standard mating sleeves. Besides that, the single-photon flux can be increased by using more efficient approaches for extraction efficiency enhancement[39]. Applying a tuneable narrow bandpass filter might further increase the flexibility of the device concept in the future. Noteworthy, our approach is independent of the material used and can be adopted to different spectral ranges. The limiting factors are the structural quality of the QD material and the spatial QD density. Also, in view of the fibre components, the excitation wavelength has to be significantly different (by at least 100 nm for the current fibre components) for a good isolation of the single-photon transition from the scattered laser light. Therefore, the strictly resonant or even quasi-resonant excitation is at this stage not possible. It would also be less practical as a very specific different wavelength of the pump would be required for each QD and only one laser source could be used for each device, which would decrease the universal character of this design, and would also increase substantially the cost of the whole system, which is now mostly dictated by the Stirling cryocooler and the excitation source. One has to keep in mind that the coherence properties in the case of the investigated structures will be in any case limited by the operation temperature (40 K) and not by the non-resonant excitation scheme. On the other hand, it is straightforward to use our concept, which can easily be transferred to other wavelengths such as the telecom C-band at 1.55 µm, with optical above-bandgap and wetting-layer excitation as well as electrically triggered structures where the pump rejection is not required at all, so it depends only on the availability of suitable QD emitters. The purity of the single photons is determined by the QD material itself. Thus, it is a matter of choosing a properly isolated transition with low emission background, which is not a limitation of the implemented concept itself, but relies more on the development of high-quality QD material at telecom wavelengths[25].

**CONCLUSIONS**

We demonstrated a user-friendly fully fibre-coupled triggered source of single-photons in the telecom O-band suitable for applications in long-range quantum communication schemes. The single photons are emitted by a semiconductor QD, deterministically integrated into a micromesa and on-chip coupled to a high NA customized single-mode fibre. The QD sample is cooled by a compact Stirling cryocooler at 40 K. The main ingredient of the proposed solution are: specially designed fibre components, the deterministic in-situ fabrication of mesa structures following the numerically obtained structure design (100% yield and spatial accuracy better than 40 nm) around the target QD and ultra-precise interferometric method for fibre alignment (accuracy below 50 nm) with respect to the mesa centre. Combining these developments resulted in a device performance with a probability of multiphoton events as low as 15% and the maximal single-photon generation rate at the single-mode fibre output of 73 kHz. The long-term stability of the optical output of the stand-alone SPS is better than 1.5% (standard deviation). Our user-friendly device concept does not require a supply of cryogenic liquids, is robust and provides a hardware solution being compact, mechanically stable and portable. It does not require any additional adjustments or post selection of the single photons by the user as the filtering fibre systems are already integrated. Therefore, the end user can operate the source in a plug&play fashion, as at the output it has a standard telecommunication single mode fibre which delivers the train of triggered single photons at 1294.7 nm for the used QD. It is worth mentioning in the context of the obtained results that similar probability of multiphoton events has been proven to be sufficient for the realization of QKD over 35 km and it outperforms the laser-based approach[40]. Additionally, the possibility of tuning the QD-based SPSs with external strain and static electric field[41,42] as well as electrical excitation[43] could be easily integrated in our source, rendering its application potential even larger. Therefore, these results pave the way to real-word application of QD-based fibre quantum networks.

**METHODS**

**Sample growth**

The QDs were formed by self-organization during metalorganic chemical vapour deposition (MOCVD) in the Stranski-Krastanow growth mode. Starting with a GaAs wafer, first the bottom DBR with 15 pairs of GaAs/Al$_{0.9}$Ga$_{0.1}$As layers with 98.3/113.8 nm thickness on a 300 nm GaAs buffer was grown at 700 °C followed by a 505.4 nm thick additional GaAs layer. For the growth of the QD layer the temperature was decreased down to 500 °C. The active region is constituted of InGaAs QDs (formed from 2.5 monolayers (ML) of InGaAs with 66% In content and flushed with 1 ML of GaAs) and is followed by a 5.5 nm thick InGaAs strain reducing layer (SRL) with an In gradient from 30% at the bottom down to 10% at the top. After initial capping with 2 nm of GaAs the capping layer consists of 612.7 nm of GaAs grown at 615 °C.

### In-situ electron-beam lithography

The applied fibre positioning method[33,44] requires that a single QD is located with high accuracy in the centre of a micrometer-sized nanophotonic structure. For that purpose, in-situ electron-beam lithography[45] optimized for 1.3 µm emission wavelength and with an overall positioning accuracy below 50 nm[34] was utilized. In this procedure $(310 \pm 2)$ nm of CSAR e-beam resist (measured by ellipsometry) diluted to a solid content of 6.5% was spin coated on the sample surface. Next, the sample is mounted in the in-situ EBL system and cooled down to cryogenic temperatures (15 K). At first the cathodoluminescence mapping was performed with 40 ms integration time for each pixel to identify the most suitable QDs for further processing. The criteria at this step are the spatial isolation of the QD, the emission intensity and the spectral range of emission. In this context it is important to note that QD emission needs to fit within the bandwidth of the fibre bandpass filters which span the range of 1293 - 1295 nm. The mapping dose must be below the onset dose for inverting the resist and in this particular processing 8 mC/cm$^2$ was used. After identifying a suitable QD the lithography step is performed still at cryogenic temperatures within the same in-situ EBL system. The resist is exposed using 25 mC/cm$^2$ electron dose and single cylindrical mesa structures are patterned in each writing field. The nominal diameters of the mesas on this sample are: 1050, 1075 and 1090 nm, and a mesa height (corresponding to etching depth) equals to $(620 \pm 5)$ nm. Further processing was performed at room temperature in the cleanroom. It includes resist development and dry etching (reactive ion etching) of the patterned structures. Afterwards, scanning electron microscopy (SEM) in top view configuration (no tilt angle) was performed to determine the actual mesa diameters and the etching depth was verified via profilometer measurements.

### Positioning of the fibre with respect to the mesa

For positioning of the optimized high-NA single mode fibre with respect to the mesa centre prior to gluing a zirconia ferrule with the fibre to the sample surface, an interferometric method detailed in[33,44] was used. The alignment procedure is performed at room temperature which, importantly, does not rely on the actual QD signal (which is not detectable at room temperature), but takes advantage of the deterministic character of the mesa structure fabrication with a single QD in the centre. The position of the fibre is adjusted for the centre of the mesa based on measurements of the topography of the sample utilizing the interference between the spectrally broad signal form the supercontinuum reflected from the fibre facet and the surface of the sample dependent on the distance between the fibre and the sample surface. Both the fibre and the sample surface are first positioned horizontally using piezo actuators and the fibre is further moved across the sample surface at constant distance. The analysis of the spectral interference fringes allows us to position the fibre with respect to the mesa centre with 50 nm accuracy (for microlenses with diameter smaller than 2 µm), i.e., with a deviation much smaller than the diameter (2.5 µm) of the single mode fibre core. After having aligned the fibre with respect to the mesa centre leaving an air gap between mesa and fibre end facet of about 0.5 µm, the fibre is set in physical contact with the sample surface which is crucial for long-term mechanical and thermal stability. Then the fibre ferrule is glued to the sample surface with ceramic UV-cured glue exhibiting a low thermal expansion coefficient of only 14 ppm/°C (compare to the GaAs coefficient of 5.73·ppm/°C). It is important to note that this glue is not transparent in the spectral range of interest, so it is only applied outside the ferrule leaving the fibre end facet and the mesa top surface free of any glue.

**Fibre-components**

The fibre components specially designed for the presented single-photon source include: a single-mode fibre with high (0.42) numerical aperture, broad bandpass filter to remove/suppress unwanted spontaneous emission background of the pulsed laser source, a fibre filter responsible for delivering the optical excitation to the sample (reflection channel) as well as spectrally suppressing it out from the detection path to provide high transmission for the actual QD signal (transmission channel), and finally a narrow bandpass filter (0.6 nm) for selecting the target emission line of the fibre-coupled QD. Additionally, the customized fibre coupled to the QD has to be spliced with a standard telecom fibre to provide an easy to integrate output of the device. The transmission of each component was evaluated with respect to the reference measurement in which a tested component was exchanged with a simple patch cord to account for spectral characteristics of the source and the detection system. Additionally, the noise level of the OSA itself was measured each time for illustrating the dynamic range of the detector to verify whether it is large enough or the result of the measurements constitute only a lower limit of the isolation provided by the respective fibre element.

The customized fibre was optimized in terms of numerical aperture and residual thermal stress for safe operation at cryogenic temperatures. The high numerical aperture (NA = 0.42) is achieved by using highly Ge doped core with 40% mol of $GeO_2$. Such a high level of doping results in a huge difference of thermal expansion coefficients between the core and the cladding of the fibre which might lead to fibre breaking either during fabrication of the fibre or cooling it down for low-temperature measurements of the QD signal. To reduce the stress level, a fibre with a three-step doping profile (40, 13 and 5% mol), resulting in the similar refractive index profile (1:2:3 diameter ratios), was fabricated following the design obtained via the finite-element method simulations. This approach resulted in a numerical aperture of 0.42 with a cut-off wavelength of 1050 nm for a fibre with 2.5 μm core diameter. The customized fibre is terminated with a zirconia ferrule polished into the spherical end-face. It is glued directly to the QD-micromesa and further sealed by an epoxy glue in a specially designed vacuum feedthrough to a portable Stirling cryocooler. The customized fibre has to be further spliced to a standard telecom single mode fibre (core diameter of 9 μm) and the main challenge here is to overcome the 3.6 factor between the fibre core diameters. A low-loss splice (0.2 dB in both directions) has been achieved using glass processing station with $CO_2$ laser splicer via the thermal core expansion technique which relies on equalizing the fibre core diameters by controlled heating of the splice area causing the $GeO_2$ dopant from the core to diffuse to the cladding, eventually creating a gradual low loss splice.

All filters were fabricated in the thin film filter (TFF) technology. The bandpass filters have a bandwidth of 0.6 nm with a flat-top characteristic which is important to maximize the emission line transmission through the filter. It is also possible to fabricate filters with a different bandwidth down to 0.3 nm (at the expense of higher loss of 1.2 dB). For their fabrication, a commercially available TFF CHIP was used and the central wavelength of the actual filter can be tuned by 2 nm via the angle of incidence on the chip which can be tuned on the fabrication stage of the filter. The tuning range of the central wavelength is limited by the change in the shape of the bandwidth and polarization selectivity appearing for high angles of incidence to 2 nm at maximum and 1 nm in practice. The parameters of the fibre filters are summed up in Table 1.

*Table 1. Parameters of the specialty fibre filters.*

| Parameter | Pumping filter | | Narrow bandpass filter |
|---|---|---|---|
| | Transmission channel | Reflection channel | |
| Maximal insertion loss [dB] | 0.70 | 4.20 (SM) <br> 1.80 (MM) | 0.50 |
| Bandwidth [nm] | 1150 ÷ 1600 | 785 ÷ 1000 | central wavelength +/- 0.3 |
| Laser attenuation [dB] | > 40 (per chip) | | > 2.50 |

The characteristics of the fibre filters (see Fig. 2) were measured using fibre-coupled supercontinuum light source (NKT Photonics SuperK Versa) and an optical spectrum analyser OSA (Yokogawa AQ6370B) covering the spectral range of 600 nm to 1700 nm. The transmission of the filters was determined as a difference between the transmission of setup with the filter and a reference measurement in which the filter was replaced by a fibre patch cord. Additionally, the noise level of the OSA was measured showing in which cases the measurement results (in particular signal attenuation by the filter) are limited by the sensitivity of the detection system itself. In this case only the lower limit for the attenuation can be determined due to limited dynamic range of the detector.

**Experimental setup – spectroscopy measurements**

In general, two experimental configurations were used for the spectroscopy study of the SPS. The common part of the two configurations was the fibre-glued sample mounted in the Stirling cryocooler (base temperature 38 K) with optical excitation (laser output filtered with the broad bandpass filter) delivered via the customized fibre pumping filter. The non-resonant excitation of the investigated QD structures was realized using an electrically-triggered fibre-coupled semiconductor diode laser custom-designed by PicoQuant. The laser is built around a pre-selected laser diode emitting at 805 nm. Special driving electronics permit freely selectable repetition rates up to 100 MHz using internal or even external triggering. The laser emits pulses with a pulse width <50 ps (FWHM) with average power of a few milliwatt, which can be further reduced by a built-in computer-controlled attenuator allowing to adjust the excitation power to the requirements of the QD structures. The difference in the two experimental configurations appears in the way the signal from the sample was filtered spectrally. For the pre-characterisation of emission from the sample, the optical signal was out-coupled from the output port of the fibre filter for pumping to free-space and filtered spectrally via a 0.32 m focal length spectrometer with 600 groves/mm grating blazed at 1000 nm providing 0.4 nm bandwidth at its output. The PL signal was further coupled to the single-mode fibre connected to a single-photon counting module (superconducting NbN nanowire detector with 20% quantum efficiency - SNSPD). This configuration was used to measure photoluminescence (PL) spectra of the QD in a broad spectral range to identify proper excitation conditions for autocorrelation measurements and in particular to select a proper bandpass filter for the second, all-fibre configuration. For the measurements on the actual fully fibre-coupled device, the output of the pumping filter was connected to the narrow bandpass filter and further via the output connector of our stand-alone SPS to a 50:50 beam splitter based on SM fibres. Each output of the beam splitter was then connected to a SNSPD for autocorrelation measurements which were carried out using a multichannel picosecond event timer (PicoHarp300) with 256 ps time-bin width. The measured histograms were fitted with the following function[17]:

$$g_{fit}^{(2)}(\tau) = g_{bg} + g_{auto}f(|\tau|) + \sum_{n\neq 0}f(|\tau - nT_0|),$$

where $g_{bg}$ corresponds to the background counts, $g_{auto}$ indicates our figure of merit – $g^{(2)}(0)$ and f(τ) is the normalized bi-exponential function with $T_0$ is the distance between the consecutive pulses (corresponding to the repetition rate of the excitation laser) which describe the auto-correlation between the photons emitted within different pulses. The peak height for the non-zero delay peaks determined from the fitting procedure was further used to normalize the coincidence events histogram to obtain the time delay-dependent second order correlation function $g^{(2)}(\tau)$.

**ACKNOWLEDGEMENTS**

This work was funded by the FI-SEQUR project jointly financed by the European Regional Development Fund (EFRE) of the European Union in the framework of the programme to promote research, innovation and technologies (Pro FIT) in Germany, and the National Centre for Research and Development in Poland within the 2nd Poland-Berlin Photonics Programme, grant No. 2/POLBER-2/2016 (project value 2 089 498 PLN). Support from the German Science Foundation via CRC 787 and the Polish National Agency for Academic Exchange is also acknowledged.

**AUTHOR CONTRIBUTIONS**

AM, GS and SR initiated the research. AM wrote the manuscript with input from all the co-authors and supervised the optical measurements of the device; KŻ performed the measurements of the characteristics of the fibre components as well as customized fibre positioning and gluing to the semiconductor structure under supervision of WU who designed the fibre feedthrough for the Stirling cryocooler and coordinated the work related with fibre transmission system for the device; NS performed the CL measurements, the patterning of the sample as well as its structural characterization under supervision of SRo; OK prepared the figures, performed the correlation measurements and the stability tests of the device; JG grew the QD sample under supervision of SR; JO performed the finite element method simulation for the customized fibre design; KP and GW fabricated the customized fibre under the supervision of PM; KD and MD fabricated and characterized the fibre components under supervision of MDł; SR and GS co-coordinated the entire project, and were directly responsible for coordinating of the work regarding the QD fabrication and spectroscopy, respectively, as well as interpretation of obtained results and proposing the idea and approach for realization of the device; KL and AB developed and optimized the laser source for optical excitation of the single-photon source; P-IS, LZ and SB performed numerical modelling of the QD-mesa structure design.

**COMPETING INTERESTS STATEMENT**

The authors declare no competing interest.

three-dimensional in situ electron-beam lithography. *Nat. Commun.* **6**, 7662 (2015).

**FIGURE LEGENDS**

Fig. 1. a) Scheme of the fully fibre-coupled single-photon source - the frame marks the device with a standard telecom single-mode FC/PC fibre connector as output. b) Image of the actual device with Stirling cryocooler on the left, pulsed excitation laser on the bottom-right and the fibre components secured on the back wall. The ventilation openings in the housing and extra fans provide air flow for cooling of the Stirling cryocooler, which is operated in vertical position and mounted to the metal frame. c) Active region and sample patterning: spatial low-temperature (15 K) cathodoluminescence map in the spectral range corresponding to the target wavelength (1293-1295) nm with the emission intensity color-coded and the target QD marked by a black circle. Inset: scanning electron microscopy image of the patterned sample with mesa etched at the position of the target QD marked by a black circle.

Fig. 2. Spectral characteristics of the fibre components measured with the supercontinuum and optical spectrum analyser (OSA): a) Broadband bandpass filter for the excitation laser. b) Single-mode (SM) pumping filter – reflection channel through which the optical excitation is delivered to the sample after cleaning up its spectrum with the filter presented in a). c) SM pumping filter – transmission channel through which the QD signal is delivered to the detection system. The laser radiation is blocked by -40 dB; d) SM narrow bandpass flat-top filter (0.6 nm full width at half maximum) with a central wavelength of 1294.65 nm; inset: set of fiber-based narrow bandpass differing in the adjustable central wavelength covering 2 nm spectral range of 1293 to 1295 nm. The orange curves show the transmission of the investigated component with respect to the reference signal level and the blue one depicts the OSA noise level.

Fig 3. Optical properties of the stand-alone SPS: a) Normalized coincidences histogram measured under pulsed non-resonant excitation (0.75 µW average power at 80 MHz) in all-fibre configuration at T = 40 K (black curve) together with a fitting curve (red); inset: corresponding spectrum measured in all-fibre configuration including the fibre-based narrow bandpass filter (its bandwidth is marked with the dashed vertical lines); b) $g^{(2)}(0)$ (left scale, blue symbols) and single photon flux rate at the output of the demonstrator (right scale, red symbols), measured under pulsed non-resonant excitation at T = 40 K as a function of average excitation power; the red dashed line marks the $g^{(2)}(0) = 0.5$ defining the limiting value for single-photon operation; the error bars were obtained as the sum of the errors of fitting parameters; c) Stability test of the SPS: histogram of the count rates (averaged over 10 min) on one of the SNSPDs measured for 18 hours. A statistical analysis yields a mean value of 1151 counts per second (cps) and a standard deviation of 16 cps which corresponds to a 0.014 relative standard deviation.